\documentclass[12pt,preprint]{aastex}

\slugcomment{}

\shorttitle{Broad Line Region in NGC 3998}
\shortauthors{Devereux et al.}

\begin{document}


\title{STIS Spectroscopy of the Central 14 pc of NGC 3998: Evidence for an Inflow.}


\author{Nick Devereux}
\affil{Department of Physics, Embry-Riddle Aeronautical University,
    Prescott, AZ 86301}
\email{devereux@erau.edu}



\begin{abstract}

Prior imaging of the lenticular galaxy, NGC 3998, with the ${\it Hubble~Space~Telescope~(HST)}$ revealed a small, highly inclined, nuclear ionized gas disk, the kinematics of which indicate the presence of a 270 million solar mass black hole. Plausible kinematic models are used to constrain 
the size of the broad emission line region (BELR) in NGC 3998 by modeling the shape of the 
broad H${\alpha}$, H${\beta}$ and H${\gamma}$ emission line profiles. The analysis indicates
that the BELR is large with an outer radius  ${\sim}$ 7 pc, 
regardless of whether the kinematic model is represented by an accretion disk or a spherically symmetric inflow. The electron temperature in the BELR is ${\le}$ 28,800 K consistent with photoionization by the AGN. 
Indeed, the AGN is able to sustain the ionization of the BELR, albeit with a high covering factor ranging between 20\% and 100\% depending on the
spectral energy distribution adopted for the AGN. The high covering factor favors a spherical distribution for the gas as opposed to a thin disk. If the gas density is
${\ge}$ 7 ${\times}$ 10$^{3}$  cm${^{-3}}$ as indicated by the broad forbidden [S II] emission line ratio, then interpreting the broad H${\alpha}$ emission line in terms of a steady state spherically symmetric inflow 
leads to a rate ${\le}$ 6.5 ${\times}$ 10$^{-2}$ M${_{\sun}}$/yr which exceeds the inflow requirement to explain the X-ray luminosity 
in terms of a radiatively inefficient inflow by a factor of ${\le}$18.
 \end{abstract}





\section{Introduction}

NGC 3998 is a lenticular galaxy located at a distance of 14.13 Mpc \citep{Ton01}. Prior imaging with the ${\it Hubble~Space~Telescope}$ (${\it HST}$) 
revealed a small (${\sim}$ 3${\arcsec}$) highly inclined nuclear ionized gas disk \citep{Pog00}
the kinematics of which indicate the presence of a 2.7$^{+1.5}_{-1.0}$ ${\times}$ 10${^{8}}$ M${_{\sun}}$ supermassive black hole (BH) \citep{DeF06}. NGC 3998 also hosts a time variable, compact core, flat-spectrum radio source \citep{Wro84} and two extended regions, described as double lobes by \citet{Fil02}.  The nucleus of NGC 3998 is also 
variable in the ultra-violet (UV) as summarized by \cite{Mao07} and in X-rays \citep{Pia10}. \cite{Era10a} have compiled the radio to X-ray spectral energy distribution revealing that the nucleus of NGC 3998 is radiating at 4 x 10${^{-4}}$ of the Eddington luminosity limit. Spectroscopically, the nucleus of NGC 3998 is classified as a LINER, with broad H${\alpha}$ and H${\beta}$ emission lines \citep[][and references therein]{Ho97} and a broad Mg II ${\lambda}$2800 emission line \citep{Rei92} which did not vary during the  3 year span of the ${\it International~Ultraviolet~Explorer~(IUE)}$ observations. Little is known about the time variability, or lack thereof, for the broad emission lines seen in the visible. 

NGC 3998 has been a frequent target for the ${\it Hubble~Space~Telescope~(HST)}$. Visible and UV spectra have been obtained with the ${\it 
Space~Telescope~Imaging~Spectrograph~(STIS)}$ at three different epochs spanning 5 years, from 1997 to 2002, allowing an investigation of whether or not the broad emission lines (BELs) respond to the aforementioned continuum variations. Additionally, the ${\it HST}$ spectroscopy was obtained with considerably higher angular and spectral resolution than the ${\it IUE}$ observations. 
Thus, the ${\it HST}$ observations provide a renewed opportunity to investigate the origin of the UV continuum and
the excitation of the associated emission lines for an unresolved but smaller ${\sim}$ 14 pc diameter region centered on the active galactic nucleus (AGN).

NGC 3998 is one of an important subset of single peaked broad line AGNs whose central mass has been measured \citep{DeF06}. Since the broad line region (BLR) is unresolved, the spatial distribution of emission cannot be directly measured.
However, the three dimensional gravitational field strength is known, thus the relationship
between velocity and radius may be established, given a kinematic model for the broad emission line gas.
In this way, one can, in principle, exploit the exquisite velocity resolution of ${STIS}$ to model the broad emission line profiles and determine a size for the BLR that heretofore has previously only been possible using reverberation mapping techniques \citep{Pet93,Pet01}.
Profile fitting complements reverberation mapping as the latter yields BLR sizes for a different sample of more
luminous AGNs.
NGC 3398 radiates well below the Eddington luminosity limit and is unable to drive an outflow thereby simplifying the interpretation of the broad emission lines by eliminating one of the possible explanations for their origin. The wider context for this investigation is to better understand the origin of single peaked broad lines, as these are by far the most ubiquitous type of broad line associated with AGNs \citep{Str03}.

The layout of the paper is as follows. In section 3, the BELs seen in
NGC 3998 are evaluated in the context of
the various models proposed for their origin, namely inflow, outflow, rotating gas disks 
and the atmospheres of stars illuminated by the 
AGN. Viable models are discussed in section 4 with the conclusion 
following in section 5 that inflow provides the best explanation for the BELs in NGC 3998. 
We begin, however, with section 2 and a review of the emission lines observed in the nucleus of NGC 3998.

\section{NGC 3998 Emission Lines }


NGC 3998 has been observed with ${\it STIS}$  twice using the G750M grating, twice using the G430L grating and one time using each of the G230L and
the G140L gratings. Table 1 summarizes the observations and the resulting spectra are presented in Fig 1. 
A ${\it STIS}$ spectrum showing the broad H${\alpha}$ emission line has been presented previously
by \cite{DeF06} based on observations obtained under PID 7354. 
\cite{GD04} presented the G430L spectrum obtained under PID 8839. However, the G130L and G230L spectra
are shown for the first time in Fig. 1. 

Multiple exposures obtained using the same grating and with the same integration time have been combined using the ${\bf STSDAS}$ task ${\bf ocrreject}$.
Dithered exposures were shifted using the ${\bf STSDAS}$ task ${\bf sshift}$ prior to combining. Subsequently, the ${\bf STSDAS}$ task ${\bf x1d}$ was used to perform a 7 pixel wide extraction along the slit direction and centered on the nucleus. Each extraction samples ${\ge}$ 80\% of the encircled energy for an unresolved point source \citep{Pro10}.

The G750M spectra reveal a broad H${\alpha}$ line with a FWHM ${\sim}$ 2000 km/s.  Another broad line
appears at ${\sim}$ 6650{\AA} but this is actually a blend of two [SII] lines. The two sets of G750M spectra are virtually identical even though the
observations are separated by four years.
The G430L spectrum reveals a wide swath of emission lines and is virtually identical to the one obtained two years later. 
Collectively, the G750M and G430L spectra resolve the H${\alpha}$, H${\beta}$ and H${\gamma}$ lines.
The G240L spectrum reveals the broad Mg II ${\lambda}$2800 and the bright C II] ${\lambda}$2326 line reported previously
by \cite{Rei92}. However, the higher signal to noise ratio and spectral resolution of the modern spectra reveal many more emission lines on a flat continuum that extends into the G140L spectrum which also shows a broad He II ${\lambda}$1650 emission
line, strongly absorbed C IV ${\lambda}$1549 and Ly${\alpha}$ emission and several other absorption lines.
A more detailed description of the spectra follows, beginning with the Hydrogen (H) Balmer lines.

\subsubsection{Broad H${\alpha}$ Emission }

The broad H${\alpha}$ emission line profile is featureless and is virtually identical in the two separate G750M observations.  There is no direct evidence for the [N II] vacuum wavelength 6549.85 and 6585.28 {\AA} emission lines in a 7 pixel extraction centered on the nucleus. Such lines are expected, however, if only due to foreground gas in the host galaxy, plus, they are seen 
in ${\it STIS}$ observations of other AGNs \citep[e.g.][]{NS03}. Evidently, they must be overwhelmed by the broad H${\alpha}$ emission in NGC 3398. 
Consequently, a model for the [N II] emission lines is employed which, following their subtraction, will yield a more realistic estimate for the broad H${\alpha}$ emission line flux. A systemic velocity of 1206 km/s is deduced from the peak of the H${\beta}$ emission line which can be clearly seen in the G430L spectrum
(Fig. 1). Atomic physics sets the wavelength of the fainter [N II] line relative to the brighter [N II] line,
it constrains the flux of the fainter [N II] line to be 1/3 that of the brighter [N II] line, and requires that the
[N II] lines share the same width. A single component gaussian with a width of 1000 km/s was chosen for the [N II] lines, similar 
to the model for the flanking [O I]  and [S II] emission lines, motivated by the fact that these forbidden
lines have similar widths in ${STIS}$ spectra 
of other AGNs for which the lines can be clearly seen \citep{NS03}. 
The flux for the brightest [N II] emission line has been chosen so that the difference spectrum does not show an `over-subtraction' 
of the broad H${\alpha}$ profile. The procedure is valid because the 
observed broad H${\alpha}$ emission line profile is otherwise smoothly varying. 
Since the [N II] lines are not directly observed, this procedure 
effectively sets an upper limit on the [N II] line fluxes and a lower limit to the flux of the broad H${\alpha}$ emission line. 
The flux for the ${\it narrow}$ component of the H${\alpha}$ line is poorly constrained. However, inspection of ${STIS}$ spectra 
for other AGNs reveals that the ${\it narrow}$ component of the H${\alpha}$ line is almost always fainter than the brightest
[N II] line \citep{NS03} which motivated the model shown in Fig. 2. Although subjective, the model for the superimposed
lines is vindicated as the difference spectrum reveals a broad H${\alpha}$ line profile that is very similar to the 
H${\beta}$ and H${\gamma}$ lines described in the next section.

The ${\bf STSDAS}$ contributed task ${\bf specfit}$ was used to model and subtract the forbidden lines and the results are reported
in Table 2 along with the fluxes for the two broad [S II] vacuum wavelength ${\lambda}$6718.29 and ${\lambda}$6732.67 {\AA} emission lines
and the two broad [O I] vacuum wavelength ${\lambda}$6302.04 and ${\lambda}$6365.53 {\AA} emission lines. The broad H${\alpha}$ emission
line flux deduced from the G750M spectra is ${\sim}$ 70\% larger than estimated previously by \cite{Ho97} using ground
based observations with good agreement on the broad line width. 

Figure 2 illustrates that the consequence of subtracting the forbidden emission line model is to reveal a broad
H${\alpha}$ emission line symmetric about the ${\lambda}$6591{\AA} wavelength expected for the systemic velocity of 1206 km/s. 
Thus, there is no apparent redshift between the 
broad H${\alpha}$ emission line and the systemic redshift of the host galaxy.

\subsubsection{Broad H${\beta}$ and H${\gamma}$ Emission }

The H${\beta}$ (Fig. 3) and H${\gamma}$ (Fig. 4) emission lines are virtually identical in the two separate observations obtained
using the G430L grating. They are compromised only slightly by [O III] emission lines
which were modeled and subtracted using the
${\bf STSDAS}$ contributed task ${\bf specfit}$ as illustrated in Fig. 3 and Fig. 4. The [O III] vacuum wavelength ${\lambda}$4364.44 {\AA}
deserved special attention as it is very faint and unresolved.  A model was employed
for the [O III] ${\lambda}$4364.44 line in which the wavelength and line width was fixed 
and the flux adjusted so as to not over-subtract the H${\gamma}$ line in the difference spectrum (Fig. 4). This yielded the upper limit for the [O III] ${\lambda}$ 4364.44 line reported in Table 3.  Subtracting the [O III] lines reveals broad H${\beta}$ and H${\gamma}$ emission lines, both of which are symmetric about the central wavelengths expected for a systemic velocity of 1206 km/s.
Unfortunately the analysis can not reliably be extended to include other Balmer lines as they are simply too faint and
even more confused with emission from other ions.
Consequently, emission line fluxes are reported in Table 3 for all the lines that can be reliably resolved and measured in the G430L spectrum
including 
H${\beta}$, H${\gamma}$, the blend of the vacuum wavelength ${\lambda\lambda}$3727.09, 3729.88 {\AA} [O II] lines,  the two [O III] vacuum wavelength ${\lambda}$4960.30 {\AA} and ${\lambda}$5008.24 {\AA} emission lines, and the 
upper limit for the vacuum wavelength ${\lambda}$4364.44 {\AA}  [O III] emission line. 

\subsubsection{Similar Balmer Line Profiles}

Figure 5 illustrates the striking similarity between the three broad H Balmer lines in NGC 3998 when they are normalized to their respective peak intensities and the wavelength scales converted 
to velocity using the non-relativistic Doppler equation. The similarity between the emission line profiles is particularly impressive considering the variety of models employed to subtract the superimposed forbidden lines.

The Balmer line profiles are similar to those seen in M81 \citep{Dev07} if
slightly narrower. Such single peaked line profiles may be
produced by radiation from a spherically symmetric shell of gas in radial motion as noted 
previously by \citet{Dev07}. Single peaked emission line profiles may also be produced
by accretion disks \citep{Che89, Era01}, bipolar flows \citep{Zhe90} and the
atmospheres of stars orbiting close to the AGN \citep{sco88, Ale94, ale97}. Each of these models will
be discussed further in section 4.

\subsubsection{Balmer Decrements}

The observed Balmer decrements, H${\alpha}$/H${\beta}$ = 3.43 ${\pm}$ 0.05 and H${\beta}$/H${\gamma}$ = 2.83 ${\pm}$ 0.11,  are
significantly different from the Case B values, 2.75 and 2.1, respectively
in the sense that the observed values are systematically 25\% and 35\% higher, respectively. Interpreting these ratios in terms of dust extinction leads to a color excess $E(B - V)$ ${\sim}$ 0.2 mag and  ${\sim}$ 0.7 mag respectively, values that are inconsistent with each other and higher than the ${\sim}$ 0.1 mag of reddening estimated to the AGN X-ray emission by \cite{Era10a}. Such deviations from recombination theory have been noted for other LINERs \citep[e.g.][]{Bow96,Fil84} and
have been attributed to collisional excitation in gas of high density. However, the requisite high densities are not achieved
in the region producing the broad emission lines in NGC 3998 as explained further in section 4.5.

\subsubsection{C II] and Broad Mg II Emission Lines }

A very broad vacuum wavelength ${\lambda}$2798 {\AA} Mg II emission line appears in the G230L spectrum. Strong absorption prevents an accurate determination
of the central wavelength but adopting a systemic velocity of 1206 km/s for NGC 3998 leads to the conclusion that the Mg II line is asymmetric with more emission on the blue side compared to the red side. With a FWHM ${\sim}$ 6000 km/s, and a FWZI ${\sim}$ 13000 km/s, the Mg II line is by far the broadest emission line detected in NGC 3998. The Mg II emission line flux noted in Table 4 agrees reasonably well with an ${\it IUE}$ measurement reported previously by \cite{Rei92} as does the vacuum wavelength  ${\lambda}$2324 {\AA} C II] emission line flux. The C II] emission line is much narrower than the Mg II emission line, as already noted by \cite{Rei92}, but the modern spectrum shows that 
the C II] emission line is broader than the H Balmer lines and shares an asymmetry similar to the Mg II emission line.
Other emission lines are detected in the G230L spectrum, the most important of which are an unresolved pair of [O II] lines with
a vacuum wavelength of ${\lambda}$2471 {\AA}. The [O II] lines have a high critical density corresponding to ${\sim}$ 4 ${\times}$10${^6}$ cm${^{-3}}$ and provide an important data point for the line width - critical density relation described in the next section. 

\subsubsection{Broad Forbidden Emission Lines }

NGC 3998 is very unusual in that it exhibits a number of broad forbidden lines which are as broad as the H Balmer lines thus blurring the distinction between the canonical ${\it broad}$ and ${\it narrow}$ line regions in this AGN. A comparison of the line widths is illustrated in the upper panel of Fig. 6 where the FWHM has been plotted against the ionization potential for forbidden and permitted lines whose line widths could be measured.  Although
the line widths are technically {\it broad} it is a misnomer to refer to them as originating from a {\it broad line region}. This is because the forbidden line widths are comparable to the Balmer line widths which is an attribute of the {\it narrow line region}. Thus, to avoid confusion, henceforth, the emission lines will be referred to as originating from a {\it broad emission line region} (BELR), the physical properties of which are the main focus of this paper.
The [S II] and [O III] lines, in particular, will provide an important measure of the 
electron number density and temperature, respectively, for the BELR of NGC 3998 as discussed further in section 4. The lower panel of Fig. 6  illustrates that there is no correlation between the line width and critical density for the forbidden lines in NGC 3998. 
According to the diagnostic diagram of \citet[][their Fig. 5]{Kew06} 
the [O III] ${\lambda}$5008.24 /[O II]  ${\lambda\lambda}$3727.09, 3729.88   and [O I] ${\lambda}$6324.99/H${\alpha}$  emission line ratios measured with {\it STIS} qualify NGC 3998 as a Seyfert and the similarity between the permitted and forbidden line widths further qualifies NGC 3998 as a Seyfert 2.{\footnotemark}
\footnotetext{The absence of strong Fe emission lines disqualifies NGC 3998 as a Narrow Line Seyfert 1 \citep{OP85}.}

\subsubsection{Flat UV Continuum with Absorption Lines}

The G140L spectrum reveals a flat UV continuum and absorption features including N V ${\lambda}$1241, Si II ${\lambda}$1260,  Si IV ${\lambda\lambda}$1394, 1403, CII ${\lambda}$1335 and tentatively Si II ${\lambda\lambda}$ 1304,1309 and O I ${\lambda\lambda\lambda}$1302, 1304, 1306. Tentative because the absorption feature appears wider than the wavelength spread of the lines attributed to it. All of the absorption features can be identified with resonance lines and are most likely produced by absorption within the host galaxy because their central wavelengths imply a redshift that corresponds to the systemic velocity of NGC 3998. 
The flat UV continuum can be traced back to the visible through the G230L, G430L and G750M spectra.




\section{Broad Line Region Models}

\subsection{Outflow Model}

A radiatively driven outflow of gas can be ruled out for NGC 3998 because the diminutive luminosity generated by the AGN is simply unable to provide sufficient radiation pressure to overcome the gravitational force of the BH as demonstrated in the following. 

The condition for a radiatively driven wind is given by, 

\begin{equation}
\kappa L/ 4 \pi c G > M 
\end{equation}

where $M$ and $L$ are the mass and luminosity respectively, interior to a radius $r$, $c$ is the speed of light, and $G$ the gravitational constant. For NGC 3998, this condition
is not satisfied by three orders of magnitude, if one adopts the Thompson scattering opacity 
${\kappa}$ = 0.4 cm${^{2}}$/g for a pure hydrogen gas, $M$ = 
2.7 ${\times}$ 10${^{8}}$ M${_{\sun}}$ \citep{DeF06}, and a bolometric luminosity, $L$ = 3.5 ${\times}$ 10${^{9}}$ L${_{\sun}}$ \citep{Era10a}.  The disparity is simply too large to overcome even by invoking line opacities, as line driven winds are viable only for objects that radiate close to the Eddington luminosity limit \citep{King03, Mur95, shl85} whereas NGC 3998 radiates at 
substantially below that limit, by a factor of  4 ${\times}$ 10${^{-4}}$. 

A VLBI observation has revealed a jet-like structure on the northern side of the nucleus of NGC 3998 \citep{Fil02}, but this is an unlikely source for the observed
broad Balmer line emission as that outflow consists of a relativistic plasma that is ejected
essentially perpendicular to the line of sight and perpendicular to the ${STIS}$ slit orientation. Gas entrained by such jets would produce a narrow line centered at the systemic redshift which is contrary to the broad emission line that is observed.

\subsection{Inflow Model}

A steady state spherical inflow is able to reproduce the 
broad Balmer emission line profiles 
observed for NGC 3998 and with the minimum of 
assumptions. 

The relationship between velocity and radius is determined by the mass distribution, 
\textit{M(r)},
to be

\begin{equation}
v(r) = -\sqrt{2 G M(r) / r}
\end{equation}

where \textit{M(r)} includes a point mass, ${M_{\bullet}}$, representing the
 black
hole, embedded in the center of an extended star cluster. In a spherical coordinate system, the observed radial component of the velocity for a particle at position $(r,\theta,\phi)$, is given by

\begin{equation}
v_{o} = v(r) cos \theta cos \phi
\end{equation}

where -${\pi}$/2~${\le}~{\theta}~{\le}~{\pi}$/2 and 0~${\le}~{\phi}~{\le}~2{\pi}$

As illustrated in Fig. 5, a broad line profile, ${\Phi(v)}$, with the characteristics of the one observed in NGC 3998 may be produced by
generating a histogram of observed velocities $\{v_{o,i}\}$ for a system of particles indexed by $i$ and distributed randomly within a 
spherical volume 
\begin{equation}
\Phi (v) =  \#\{i|v\leq v_{o,i} \le v+dv\}.
\end{equation}

Here $v_{o,i}$ denotes the observed velocity of the $i^{th}$ particle. Implicit in this model is the assumption that the particles share the same emissivity and emit isotropically.

For a steady state flow the particle number density distribution is determined by mass conservation to be

\begin{equation}
N(r) \propto r{^{-3/2}}.
\end{equation}

The distribution is used  to produce a broad line profile by creating a series of concentric spherical shells $\{S_j\}$ of radii $\{ r_j\}$ selected randomly between $r_{inner}$ and $r_{outer}$. On each
shell, $\{S_j\}$ a total of $N(r_j)$ particles is distributed randomly in position, (${\theta,\phi}$). The model differs
from Bondi flow \citep{Bon52} in that the particles are discrete and do not constitute a fluid. 

Approximately 12,000 particles are included in the model. The velocity histogram is binned by 25 km/s and subsequently smoothed using a 10 bin moving
average. The resulting model velocity resolution is comparable to that of the G430L spectra but considerably lower than the G750M spectra. The details of the
resulting line profile shape are sensitive to the range of values employed for ${\theta}$
and ${\phi}$. For example, restricting the range of ${\theta}$ such that -0.96${\pi}$/2~${\le}~{\theta}~{\le}~0.96{\pi}$/2 simulates the cavity expected to be occupied by the radio jets and causes the model profile to more closely mimic the observed profile in the region near the peak{\footnotemark}. 
\footnotetext{A feature of the inflow model is that it produces a delta function at zero velocity, a consequence of the large number of clouds moving perpendicular to the line of sight. Such a delta function is not present in the observed spectra and is therefore mitigated in the current model by invoking a cavity. }
In the inflow model, the
velocity width at the top of the line is determined by the outer radius, ${r_{outer}}$, and the width at zero intensity is determined by the inner radius, ${r_{inner}}$, and these values are largely {\it insensitive}
to the range of ${\theta}$ and ${\phi}$.
Thus, given a density distribution and a velocity law, one can use the profile shape to determine the physical size of the emitting region. Values that produce
a reasonable representation of the broad Balmer emission line profiles, shown in Figure 5, 
correspond to an
outer radius, ${r_{outer}}$ = 7 pc, and an inner radius,  ${r_{inner}}$ = 0.08 pc. The model is illustrative
and does not represent a unique solution or the result of a full exploration of the parameter space.
However, experience indicates that the model values determined for the inner and outer radius
would have to change substantially, by more than ${\sim}$ 30\%, to cause the model to 
seriously misrepresent the observed profile. Fig. 5 illustrates that the residuals between the observed profiles and the model prediction are ${\sim}$ 10\%.

\subsection{Accretion Disk Model}

Next, the kinematics of a rotating axisymmetric Keplerian disk are considered as
an explanation for the shape of the broad Balmer emission line
profiles. While there has been a novel proposal \citep{mur97} to
generate a single peak profile using a disk wind, it is not adopted
here because the AGN in NGC 3998 radiates at  4 ${\times}$ 10${^{-4}}$ of the Eddington luminosity limit. Therefore,  if an accretion disk wind is present it
is expected to be feeble and not influence the propagation of Balmer line
photons
\cite[see section 3.1 and the discussion of Arp~102B by][]{Hal96}. Instead, the simplest possible model is considered
which involves an axisymmetric, relativistic accretion disk developed
by \cite{Che89}, and adopted later by \cite{Era01} to explain the
single peak broad line profile in the LINER nucleus NGC 3065. In
addition to Doppler shifts, this model includes self-consistently all
relevant relativistic effects, such as Doppler boosting and transverse
and gravitational redshifts.

The axisymmetric disk model invokes five free parameters, a
dimensionless inner radius, ${\xi}_{1}$, and outer radius,
${\xi}_{2}$, an inclination angle measured from the disk normal to the line
of sight, $i$, an emissivity law of the form
$\epsilon\propto r^{-q}$, and
a velocity dispersion for the gas, ${\sigma}$ in km/s. The line profile is calculated by numerically integrating eqn. 7 of \cite{Che89} using a wavelength resolution of 2.5 {\AA} which is comparable to the G430L spectra but lower than the G750M spectra. Values of the parameters that best reproduce the Balmer line profiles observed in NGC 3998 are summarized in Table 5 and include a large outer radius corresponding to ${r_{outer}}$ = 7 pc.
The model is illustrative
and does not represent a unique solution or the result of a full exploration of the parameter space.
However, experience indicates that the model parameters
would have to change substantially, by more than ${\sim}$ 30\%, to cause the model to 
seriously misrepresent the observed profile. Fig. 5 illustrates that the residuals between the observed profiles and the model prediction are ${\sim}$ 10\%.

\section{Discussion}

Evidently, both a spherically symmetric inflow and an accretion disk are able to
reproduce the shape of the broad Balmer emission line profiles observed for NGC 3998. The mass 
distribution determines that the emitting region is large;  ${\sim}$ 7 pc (0.1${\arcsec}$) in radius for both models.
Whereas it is already known that NGC 3998 hosts a small (${\sim}$ 3${\arcsec}$) highly inclined nuclear ionized gas disk \citep{Pog00}, the kinematic evidence for a spherically symmetric inflow would represent a new phenomena for NGC 3998. Which of the two explanations is most likely to be correct
is discussed in the following.

All previous studies in which the broad Balmer emission lines have been attributed to an accretion disk have involved
line profiles whose shapes are very different and much broader than the adjacent and overlapping [N II],  [S II] and [O I] forbidden lines \cite[e.g.][]{Era03,Bar01,Shi00,Era95,Sto95, Era94}. Indeed, the distinction between the broad permitted lines and 
the much narrower forbidden lines has motivated the notion that the former are produced by gas that has a density exceeding the critical density of the forbidden lines and arises from a kinematically and physically distinct gas
component identified with an accretion disk. NGC 3998 is different. The Balmer and forbidden line widths are similar suggesting that they have a common origin. The question is do the broad emission lines originate in an accretion disk or in an inflow? The answer to this question may lie in the covering factor,
explored in more detail in the next section.

\subsection{Broad Line Region Ionization}

The ionizing continuum in NGC 3998 may be represented as a power law,

\begin{equation}
L_{\nu} = L_{o} (\nu/  \nu_o)^{-\alpha}
\end{equation}

which may be integrated to yield the number of ionizing photons, N$_{ion}$

\begin{equation}
 N_{ion} =  \int^{\nu_{max}}_{\nu_o} [L_{\nu}/ h \nu ]d\nu
\end{equation}

to yield

\begin{equation}
 N_{ion} = L_{o}  \nu_o^{\alpha} [  \nu_o^{-\alpha} -  \nu_{max}^{-\alpha}] /  (h \alpha )~photons~s^{-1}
\end{equation}

Adopting an optical to X-ray spectral index ${\alpha}$ = 1 \citep{Era10a} to represent the shape of the ionizing continuum in NGC 3998, predicts just enough ionizing photons to explain the luminosity
of the broad H${\alpha}$ line. The 
central AGN produces 7.6 x ${10^{51}}$ ionizing ph/s, integrated between 13.6 eV and 100 keV, after correcting for extinction.
For comparison, the broad  H${\alpha}$
emission line flux (Table 2) corresponds to an H${\alpha}$ luminosity,  $L(H{_{\alpha}}$), of 2.6 ${\times}$10$^6$ L${_{\sun}}$,
which, assuming 45\% of the ionizing photons are converted into H${\alpha}$ photons (Case B recombination at a temperature of 10$^4$ K) requires (7.4 $\pm$ 0.2) x ${10^{51}}$ ionizing ph/s using,

\begin{equation}
N_{ion} = L(H{_{\alpha}}){\alpha_{B}}/{\alpha^{eff}_{H\alpha}} h \nu_{H\alpha} 
\end{equation}

where ${\alpha^{eff}_{H\alpha}}$ = 1.16 x 10$^{-13}$ cm${^{3}}$ s${^{-1}}$ is the effective recombination coefficient  and 
${\alpha_{B}}$ = 2.59 x 10$^{-13}$ cm${^{3}}$ s${^{-1}}$ is the total Case B recombination coefficient.
The number of ionizing photons required to excite the 
H${\alpha}$ line, estimated using eqn. 9, is independent of the gas density and the filling factor but the comparison with the ionization available from the AGN
does imply a very high covering factor corresponding
to 97 $\pm$ 3\%. If the broad  H${\alpha}$ emission line flux is corrected for A${\rm_v}$ = 0.3 mag of extinction \citep{Era10a} then the 
number of ionizing photons required to excite the 
H${\alpha}$ line exceeds that available from the AGN{\footnotemark} \footnotetext{The ionizing deficit widens to 80\%  if the extinction is estimated from the H${\alpha}$/H${\beta}$ ratio and a factor of 7 if the extinction is estimated from the H${\beta}$/H${\gamma}$ ratio, although it is not obvious that the anomalous Balmer decrements
should be interpreted in terms of dust extinction given the similar and symmetric Balmer line profile shapes.}by 28\% causing NGC 3998 to join M81 as an example of an AGN with a BELR ${\it ionizing~deficit}$ \citep{Dev07,Ho96}.
 
That the central AGN produces 7.6 x ${10^{51}}$ ionizing ph/s, integrated between 13.6 eV and 100 keV, 
is a factor of 3 lower than an independent estimate of the ionizing photon rate by \cite{Era10b}. The discrepancy is due to different assumptions concerning the spectral energy distribution in the far-UV which is complicated by the fact 
that the nucleus of NGC 3998 is time variable.
The G140L and G230L ${\it HST}$ spectra obtained in 2002 show that ${f_\lambda}$ is approximately constant in the UV (1200 - 3000 {\AA}) at ${\sim}$ 2 x ${10^{-15}}$ erg
cm${^{-2}}$ s${^{-1}}$ {\AA}${^{-1}}$. This is consistent with the 2120 {\AA} measurement of ${\sim}$ 2.5 x ${10^{-15}}$ erg
cm${^{-2}}$ s${^{-1}}$ {\AA}${^{-1}}$ obtained in 2001 with the ${\it XMM~Newton}$ 
Optical Monitor by \cite{Pta04} and the F250W measurement of 2.2 x ${10^{-15}}$ erg
cm${^{-2}}$ s${^{-1}}$ {\AA}${^{-1}}$ obtained in 2002 with ${\it HST}$ by \cite{Mao05}. However, in 1992 the nucleus was about 5 times brighter. Then the flux measured with {\it HST} in the F175W filter was at least 1 x ${10^{-14}}$ erg
cm${^{-2}}$ s${^{-1}}$ {\AA}${^{-1}}$ \citep{Fab94}. The nucleus has apparently been declining in brightness between 1999 and the date of the most recent published UV measurement in 2003 
when the flux measured with {\it HST} in the F250W filter was 1.8 x ${10^{-15}}$ erg
cm${^{-2}}$ s${^{-1}}$ {\AA}${^{-1}}$ \citep{Mao05}. 
The ionizing photon rate estimated by \cite{Era10b} includes the high value measured by \cite{Fab94} and the higher value for the
2120 {\AA} flux favored by \cite{Pta04}
whereas the new estimate presented here does not.
As such, the rate N$\rm_{ion}$ = 7.8 x ${10^{51}}$ ionizing ph/s, represents the ionizing photon rate in the central 0.2${\arcsec}$ of NGC 3998, {\it circa} 2002. If, on the other hand, the spectral energy distribution is represented by a quasar \citep{Era10b}, then N$\rm_{ion}$ increases to 3.5 x ${10^{52}}$ ionizing ph/s, which alleviates the ionizing deficit, but still implies a high covering factor corresponding to $\sim$ 20\%.

The high covering factor $\ge$ 20\% is much easier to achieve with a spherical distribution of gas than a thin 
accretion disk and is the main argument that the broad emission lines seen in NGC 3998 are due to an inflow, an outflow being impossible for the reasons explained in section 3.1. 
The possibility of inflows has received
remarkably little attention 
in the literature as noted recently by \cite{Gas09}. Indeed, there has been nary a mention of inflows in the context of broad line profile shapes since \cite{Cap80}. Yet this most neglected of models is able to explain the broad Balmer emission lines seen in NGC 3998
and another low luminosity AGN, M81, as noted previously by \cite{Dev07}. 

\subsection{Stellar Atmospheres Illuminated by the Central AGN Model}

Another contending explanation for the broad emission lines seen in NGC 3998 is that they result from the ionization of the extended mass loss envelopes of red giant stars orbiting close to the AGN \cite[e.g.][]{sco88, ale97}. Stellar winds provide a natural explanation for the confinement and replenishment of the so called ${^{\prime}}$broad line clouds${^{\prime}}$, which is otherwise a major problem \citep{mat87}. Such a model may explain the  in NGC 3998, because the intensity of ionizing photons in the BELR at a distance, ${r}$  ${\sim}$ 5 pc, from the AGN is
sufficient to penetrate the mass loss wind to a density ${\sim}$ ${10^{4}}$ cm${^{-3}}$ which is comparable to that inferred for the BELR from
the [S II] lines (section 4.5). Following \citet{Dev07}, it can be shown 
that for a power law of spectral index ${\alpha}$ = 1, the density, ${n}$, in the wind at the penetration depth, ${d}$, of 
the ionizing photons is given by,

\begin{equation}
n(d) = 98 \times10^6 [L_{1500}/10^5 L_{\sun}]^{2/3} [r/10^{15} cm]^{-4/3} [v_W/10~km/s]^{1/3} [\dot{M}/ 10^{-5} M_{\sun}/yr]^{-1/3}  cm^{-3}
\end{equation}

Representative values for NGC 3998 are $r$ = 5 pc, $L_{1500}$ = 1.6 ${\times}$ 10${^{7}}$ L$_{\sun}$, $v{_W}$ = 10 km/s
for  the wind velocity, 
and $\dot{M}$ = 10${^{-5}}$  M${_{\sun}}$/yr \citep{sco88} and  $\dot{M}$ = 10${^{-6}}$  M${_{\sun}}$/yr for the stellar mass loss rates 
\citep{ale97}. One finds that ${n(d)}$ is 7.8 x ${10^{3}}$ cm${^{-3}}$ and 1.7 x ${10^{4}}$ cm${^{-3}}$ respectively, values that are comparable to the gas densities inferred
for the BELR from the [S II] lines discussed further in section 4.5. \citet{ale97} have explored the so called  ${^{\prime}}$bloated stars ${^{\prime}}$ (BSs) model in some detail although not for the
specific characteristics of the AGN/BEL combination in NGC 3998. Such an investigation may be worth revisiting, however, as the properties of the 
 gas in NGC 3998 may be explained in the context of a low density, optically thin, stellar mass loss wind which would allow the
production of broad forbidden emission lines with velocity widths similar to the Balmer lines. Previously, \citet{ale97} were restricted to models that suppressed
the formation of broad forbidden lines by invoking stellar mass loss winds of unrealistically high density, but it appears that the density restriction can now be lifted, at least in the
case of NGC 3998. Another attractive feature of the BSs model is that it can potentially achieve the high covering factor needed to explain the near equality between the ionization provided by the AGN and that required by the broad Balmer lines. Stellar mass loss is also the most likely origin for the inflowing gas invoked
to explain the broad Balmer line profiles.

\subsection{BELR Size}

Perhaps the most surprising result to have emerged from the analysis of the broad Balmer emission line profiles is the 
large outer radius, ${\sim}$ 7 pc, inferred for BELR of NGC 3998{\footnotemark} \footnotetext{The ${\lambda}$2798 {\AA} Mg II and ${\lambda}$2324 {\AA} C II] emission lines are broader than the Balmer lines, suggesting that they originate from a region that is closer to the AGN, but the fact that these lines are asymmetric and that the Mg II line also suffers from absorption would require more sophisticated modeling that is beyond the scope of the present paper.}corresponding to an angular diameter of 0.2${\arcsec}$.
An independent analysis of the angular size based on the encircled energy, illustrated in Fig. 7, shows that the 
BELR of NGC 3998 is spatially unresolved with the 0.1${\arcsec}$ wide slit employed for the ${\it STIS}$ G750M observations. This is because the percentage of the broad line flux measured in 1 pixel wide and 2 pixel wide extractions, as compared with the flux measured in a 7 pixel wide extraction, is consistent with that expected for a point source.
Additionally, there is no perceptible difference between the  H${\alpha}$
profile obtained with a 0.1${\arcsec}$ slit and the H${\beta}$ and H${\gamma}$ profiles
which were obtained with a 0.2${\arcsec}$ slit. However, no large difference is expected based on the modeling.
Even if the `poles' of the model spherical inflow are excluded, as would be the case when the slit width is 0.1${\arcsec}$ and slightly
smaller than the 0.2${\arcsec}$ diameter of the inflow, the model line profile shape remains unchanged. This is because
only a few points are excluded from the model. Additional evidence that the BELR is unresolved is that the 
Balmer line profiles are similar even though the observations were made with a variety of slit orientations that differed by as much as ${\sim}$ 25 degrees in position angle.

The size inferred for the BELR causes NGC 3998 to not conform to the
correlation between BLR size and UV luminosity established for quasars and high
luminosity AGNs using reverberation mapping \citep{Pet01,Pet93, Kas05}. However, as
noted by \cite{Kas05}, the correlation appears to break down
for low luminosity AGNs, which, with L(1450 ${\textrm \AA}$) = 6.3 ${\times}$ 10${^{40}}$ erg/s, estimated from the G140L spectrum,
would include NGC 3998, even allowing for the variability which is discussed in more detail in section 4.1. Nevertheless,  an extrapolation of the BLR size - luminosity relationship of \cite{Kas05} down to the low UV luminosity estimated for the AGN in NGC 3998, predicts a size for the BLR that is about 200 times smaller than the inner radius of the BELR determined for NGC 3998 by profile fitting\footnotemark \footnotetext{Compared to the size predicted using the FITEXY method of \cite{Kas05} at 1450{\AA}.}. Conversely, the large outer radius determined for the BELR of NGC 3998 using profile fitting makes it  larger than any BLR measured using reverberation mapping, 24 times larger than the previous record holder; the quasar 3C 273 \citep{Kas05}. Even though NGC 3998 is known to be variable in the UV \citep{Mao07}, no variability has been detected in the Balmer lines on a timescale of 4.4 years precluding an estimate of its reverberation based BLR size.
Of course, the BLR size - luminosity relationship of \cite{Kas05} is defined by quasars and AGNs that are orders of magnitude more luminous than NGC 3998. Thus, the very large discrepancy arising from the comparison strongly suggests that the BELR in NGC 3998 is not simply the BLR of a scaled down quasar.

\subsection{Virial Black Hole Masses}

Estimating the mass of the BH in NGC 3998 using the so called `virial method'  leads to a value that is 
substantially lower than the kinematically determined mass. For example, the formalism of \cite{Gre05}, which uses the FWHM {\it and} luminosity of the broad H${\alpha}$ emission line, underestimates the mass of the BH in NGC 3998 by a factor of 413. On the other hand, the BH in
NGC 3998 does conform to the BH mass - bulge mass correlation as noted previously by \cite{DeF06}. This dichotomy is 
regarded as further evidence that the BELR  in NGC 3998 is very different from the BLR in more luminous AGNs.

\subsection{Constraints on the BELR Gas Density and Temperature}

The fact that the two [S II]  lines are nearly as broad as the Balmer lines provides an opportunity to set a lower limit on the gas density in the BELR. 
The nominal value for the observed [S II] ${\lambda}$6742/${\lambda}$6754 
intensity ratio = 0.58 ${\pm}$ 0.24, corresponding to 
$n$  ${\sim}$ 7 ${\times}$ 10$^3$ cm$^{-3}$, but the large uncertainty permits gas densities in the range 
10$^3$ cm$^{-3}$ ${\le}$  $n$  ${\le}$ 10$^4$ cm$^{-3}$.  These densities are much lower than the 10$^9$ cm$^{-3}$ often quoted for the BLRs of more
luminous AGNs. However, the fact that the Balmer lines are wider than either of the [S II] lines may suggest that the gas density is higher than 7 ${\times}$10$^3$ cm$^{-3}$ in the region emitting the Balmer lines if the gas density increases closer to the BH. 

The limit for the observed [O III] ratio (${\lambda}$4960.30 + ${\lambda}$5008.24) / ${\lambda}$4364.44 ${\geq}$ 24 yields T ${\leq}$ 28,800 K for the electron temperature if $n$ = 7 ${\times}$10$^3$ cm$^{-3}$, which is judged to be representative of the temperature in the BELR because the vacuum wavelength [O III] ${\lambda}$4960.30 and ${\lambda}$5008.24 lines ${\it are}$ observed to be as broad as the Balmer lines. 

The ionization parameter, ${\Gamma}$, given by 

\begin{equation}
 \Gamma = N_{ion}/ 4  \pi r^2 c~n
\end{equation}

corresponds to 0.006  ${\le}$  ${\Gamma}$ ${\le}$ 48 for gas in the BELR  with 7 ${\ge}$  $r$(pc) ${\ge}$ 0.08 pc and
ionized by the central AGN. Such values for the ionization parameter are similar to those expected inside an HII region ionized
by an O5 star. That is not to say that the ionization in NGC 3998 is provided by O5 stars, but rather that the electron
temperature is expected to be similar and consistent with the limit ${\leq}$ 28,800 K  determined from the [O III] ratio (${\lambda}$4960.30 + ${\lambda}$5008.24) / ${\lambda}$4364.44.

\subsection{The Mass of Ionized Gas Required to Produce the Broad H${\alpha}$ Emission Line}

The mass of emitting gas may be deduced from the broad H${\alpha}$ emission line 
luminosity assuming standard (Case B) recombination theory;

\begin{equation}
 M_{emitting} =  L (H_\alpha) m_H /  n_H  {\alpha^{eff}_{H\alpha}} h \nu_{H\alpha} 
\end{equation}

Using an effective recombination 
coefficient ${\alpha^{eff}_{H\alpha}}$ =
8.6 x 10$^{-14}$ cm${^{3}}$ s${^{-1}}$, assuming a \textit{constant} average density $n$ ${\ge}$ 7 ${\times}$10$^3$ cm$^{-3}$,
and a luminosity $L (H{_{\alpha}}$) = 2.6 x 10${^6}$ L${_{\sun}}$ based on the broad line flux reported in Table l, leads to an upper limit on the mass of ionized gas emitting the
broad H$_{\alpha}$ line, $M_{emitting} $ ${\le}$ 4600 M${_{\sun}}$. 
If only a fraction of the gas is ionized then the upper limit on the  {\it total} (ionized + neutral) gas mass could, of course, be much higher.  Thus, the total mass of gas in the  BELR of NGC 3998 could be substantial. 

\subsection{The Filling Factor and the Inflow Rate for the BELR of NGC 3998 and an Assessment of the Inflow Scenario}
 
It is straight forward to calculate the 
filling factor, ${\epsilon}$, for a spherically symmetric inflow, once the dimensions of the emitting region have been established. 
For a uniform density medium occupying a spherical volume of radius $r$, one finds 

\begin{equation}
 \epsilon = 3 L (H_\alpha)/ 4  \pi  n_H^2  {\alpha^{eff}_{H\alpha}} h \nu_{H\alpha} r^3
\end{equation}

Again, using an effective recombination 
coefficient ${\alpha^{eff}_{H\alpha}}$ = 8.6 x 10$^{-14}$ cm${^{3}}$ s${^{-1}}$, assuming a \textit{constant} average gas density $n$ ${\ge}$ 
7 ${\times}$10$^3$ cm$^{-3}$, and a luminosity $L (H{_{\alpha}}$) = 2.6 x 10${^6}$ L${_{\sun}}$ based on the broad line flux reported in Table 1, leads to an upper limit on the  filling factor, ${ \epsilon}$ ${\le}$ 2 x 10$^{-2}$, for NGC 3998 if 
the size of the BELR, $r$ = 7 pc. The very low filling factor suggests that the inflow is not continuous but
composed of many ionized, density bounded, gas filaments. Such filaments would have approximately the same gas density and hence the same emissivity regardless of their location with respect to the central AGN, they would be optically thin and hence emit isotropically. Ionized gas filaments embrace all the essential elements of the inflow model and are commonly seen in the nuclei of active galaxies \citep{Sto09}.  
 
Having established the dimensions of the emitting region and the filling factor one 
can calculate the mass inflow rate, $\dot{m}$, for
the ionized gas, using the equation of continuity;

\begin{equation}
 \dot{m}  =  \epsilon 4 \pi r^2 v  n_H m_H
\end{equation}

The velocity at the inner radius of 1 pc is determined by the mass distribution to be 1550 km/s. Setting the gas density in the flow, ${n}$ ${\ge}$ 7 ${\times}$10$^3$ cm$^{-3}$, one obtains an
upper limit to the mass inflow rate,  $ \dot{m}$ ${\le}$ 6.5 ${\times}$ 10$^{-2}$ M${_{\sun}}$/yr. However, if only a fraction of the inflowing gas is ionized, then the upper limit on the {\it total} mass inflow rate could, of course, be higher.

The 2--10 keV X-ray luminosity adopted for the AGN in NGC 3998, $L_{2-10~keV}$, is $2.6\times 10^{41}~{\rm erg~s^{-1}}$
\citep{Era10a}. Assuming this is powered by radiatively inefficient accretion
leads to the following formula  \citep{Mer03}

\begin{equation}
 L_{2-10~ keV}= 7  \times 10^{38} M_{\bullet}^{0.97}  \dot{m}^{2.3}
\end{equation}

where $L$$\rm_{2-10~keV}$ is in ${erg~s^{-1}}$ and $M$$_{\bullet}$ is in solar masses. Under these
circumstances the accretion rate required to power the observed X-ray emission,  $\dot{m} $ $\sim 3.6 \times
10^{-3} ~{\rm M_\odot~yr^{-1}}$.

\section{Conclusions}

A new technique has yielded a size for the BELR in NGC 3998 by modeling the shape of the broad H${\alpha}$,
H${\beta}$ and H${\gamma}$ emission line profiles. The principal conclusion is that the BELR  is large, ${\sim}$ 14 pc in diameter, 
and represents an inflow, likely sustained by stellar mass loss. The large size determined for the BELR  in NGC 3998 is inconsistent with an extrapolation of the 
reverberation based BLR size - luminosity relationship. Additionally, the virial method for estimating BH masses
using the broad H${\alpha}$ emission line 
width and luminosity lead to an inconsistent BH mass for NGC 3998. Both of these relationships are
defined by the broad emission lines of quasars and high luminosity AGNs. It is therefore concluded that the 
BELR in NGC 3998 is not the BLR of a scaled down quasar but more likely identified with the narrow line region.
The AGN is able to sustain the ionization of the BELR, albeit with a high covering factor ranging between 20\% and 100\% depending on the
adopted spectral energy distribution. Such a high covering factor is most easily provided by a spherical distribution of gas as opposed to a thin disk. The electron temperature in the BELR is ${\leq}$ 28,000 K consistent with photoionization
by the AGN.
The gas density is pivotal in constraining the mass of gas in the BELR.
If the gas density is high, 
${\ge}$ 7 ${\times}$ 10$^{3}$  cm${^{-3}}$, then interpreting the broad H${\alpha}$ emission line in terms of a steady state spherically symmetric inflow 
leads to a rate ${\le}$ 6.5 ${\times}$ 10$^{-2}$ M${_{\sun}}$/yr which exceeds the requirement to explain the X-ray luminosity 
in terms of a radiatively inefficient inflow by a factor of ${\le}$18.




\acknowledgments
This research has made extensive use of the NASA Astrophysics Data System, the Atomic Line List, http://www.pa.uky.edu/~peter/newpage/
and the ${\bf STSDAS}$ task ${\bf ionic}$ for calculating the critical densities of various ions. 
Support for Program number HST-AR-11752.01-A was provided by NASA through a grant from the Space Telescope Science Institute, which is operated by the Association of Universities for
Research in Astronomy, Incorporated, under NASA contract NAS5-26555. The author thanks Michael Eracleous
for help with estimating the ionizing photon rate provided by the AGN and Ari Laor for emphasizing the subtle distinction between {\it broad lines} and {\it broad line regions}.



{\it Facilities:}  \facility{HST (STIS)}

\clearpage



\begin{figure}
\epsscale{0.8}
\begin{center}
\plotone{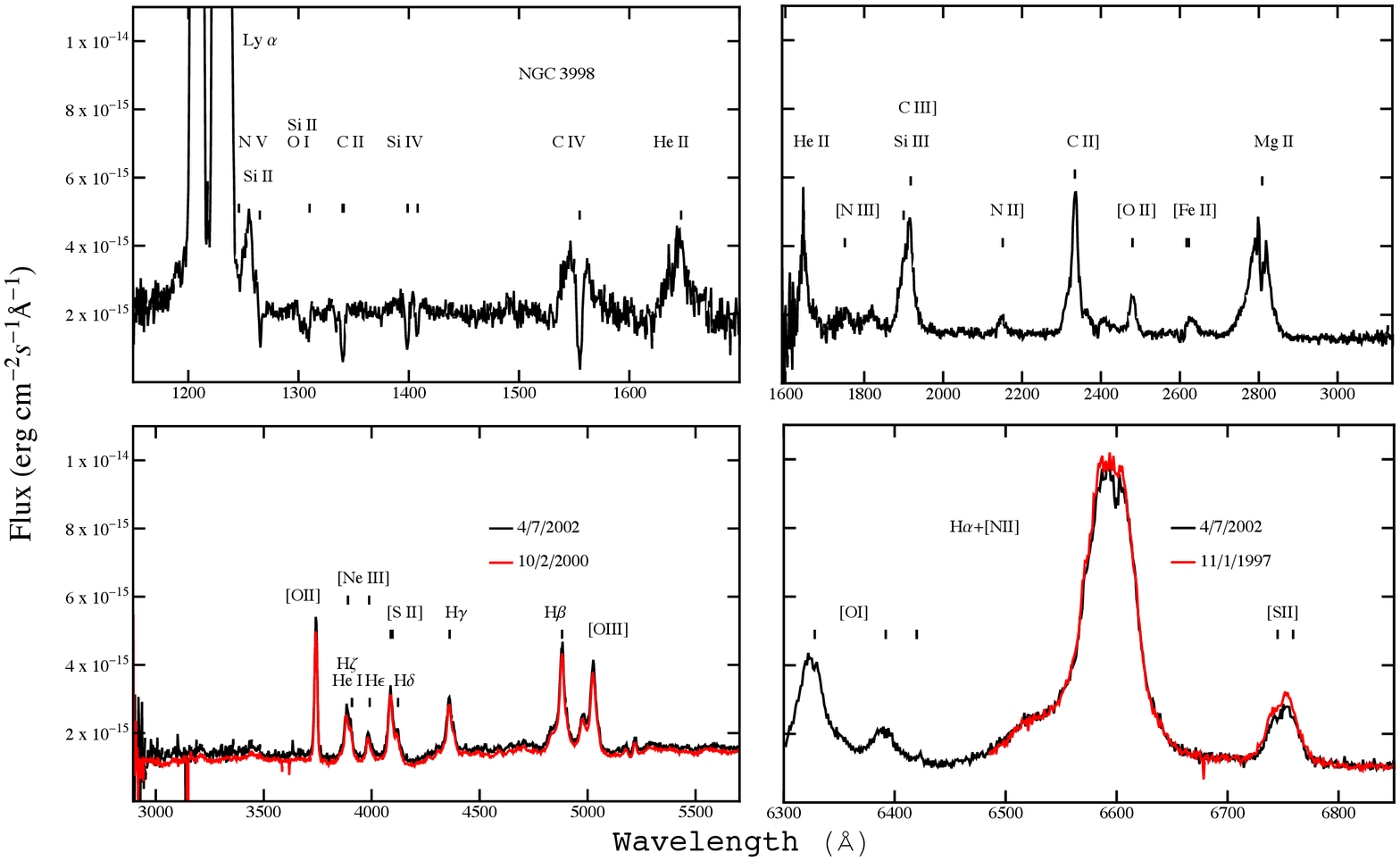}
\caption{{ Visual and UV spectra  of NGC 3998 as seen through the following gratings:  {\sl Top left panel}: G140L. {\sl Top right panel}: G230L.
{\sl Lower left panel}: G430L. Red line shows data obtained under PID 8839. {\sl Lower right panel}: G750M. Red line shows data obtained under PID 7354. Black lines for all panels show data obtained under PID 9486.}}

\label{default}
\end{center}
\end{figure}

\begin{figure}
\epsscale{0.8}
\begin{center}
\plotone{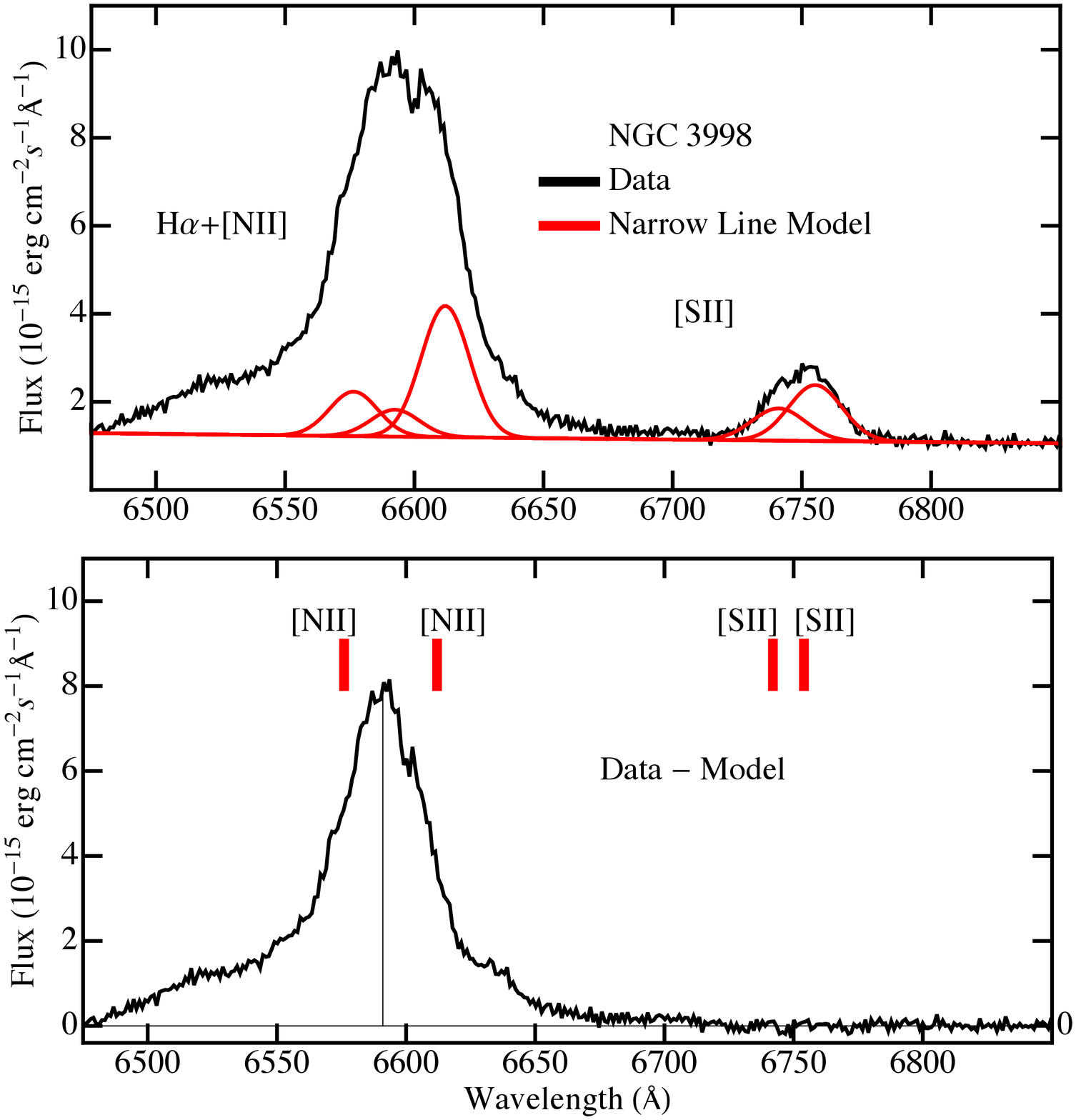}
\caption{{ Broad H${\alpha}$ emission line in NGC 3998. {\sl Top panel}: The observed spectrum
is shown in black and a model for the forbidden lines is shown in red (see also Table 2). {\sl Lower panel}: The broad 
H${\alpha}$ emission line profile
after the forbidden lines have been subtracted. The central wavelengths of the subtracted lines are
indicated in red. The vertical black line corresponds to the observed (redshifted) central wavelength of the H${\alpha}$ line }}
\label{default}
\end{center}
\end{figure}

\clearpage

\begin{figure}
\epsscale{0.8}
\begin{center}
\plotone{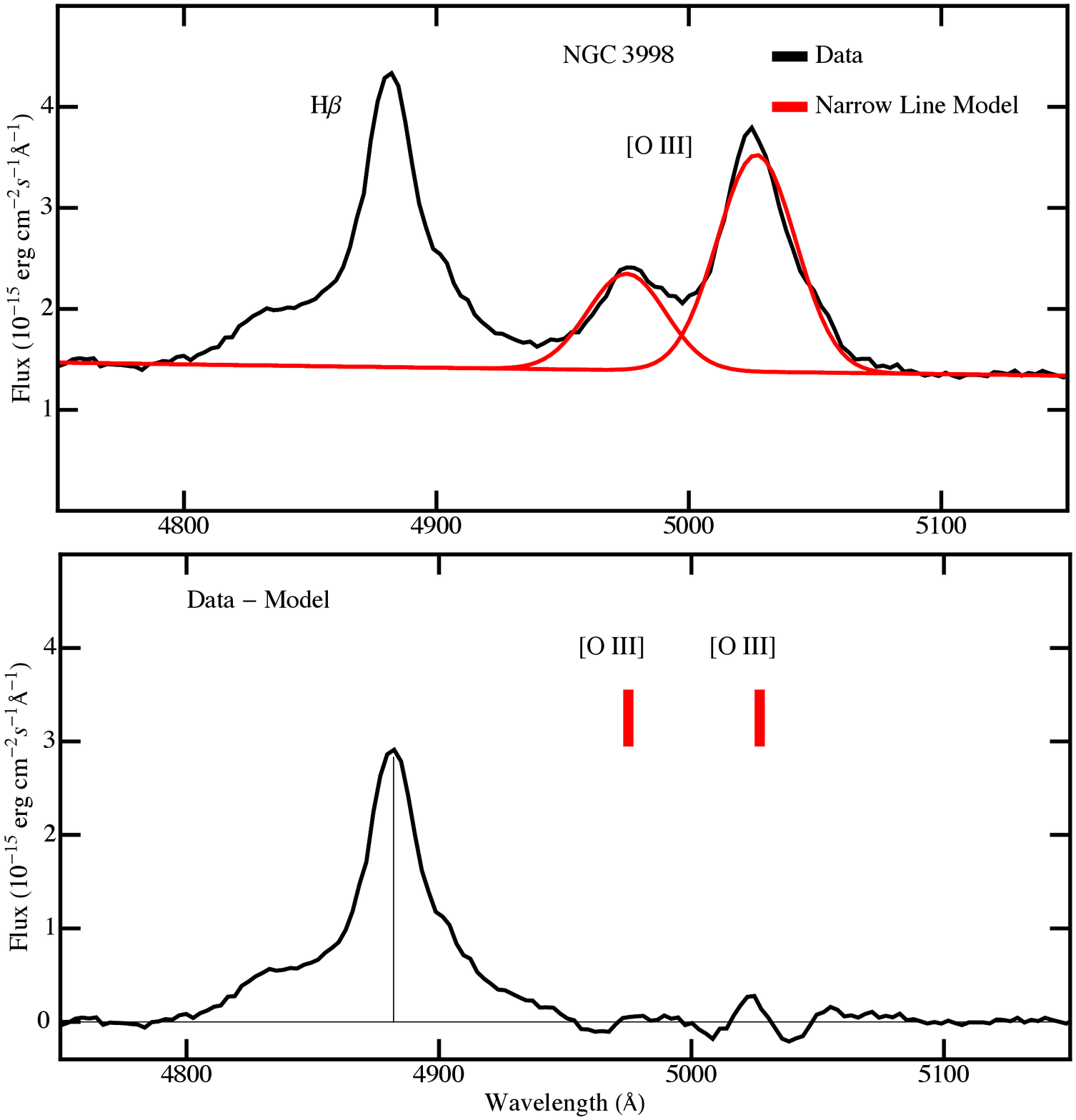}
\caption{{ Broad H${\beta}$ emission line in NGC 3998. {\sl Top panel}: The observed spectrum
is shown in black and a model for the forbidden lines is shown in red (see also Table 2). {\sl Lower panel}: The broad 
H${\beta}$ emission line profile
after the forbidden lines have been subtracted. The central wavelengths of the subtracted lines are
indicated in red. The vertical black line corresponds to the observed (redshifted) central wavelength of the H${\beta}$ line }}
\label{default}
\end{center}
\end{figure}

\clearpage

\begin{figure}
\epsscale{0.8}
\begin{center}
\plotone{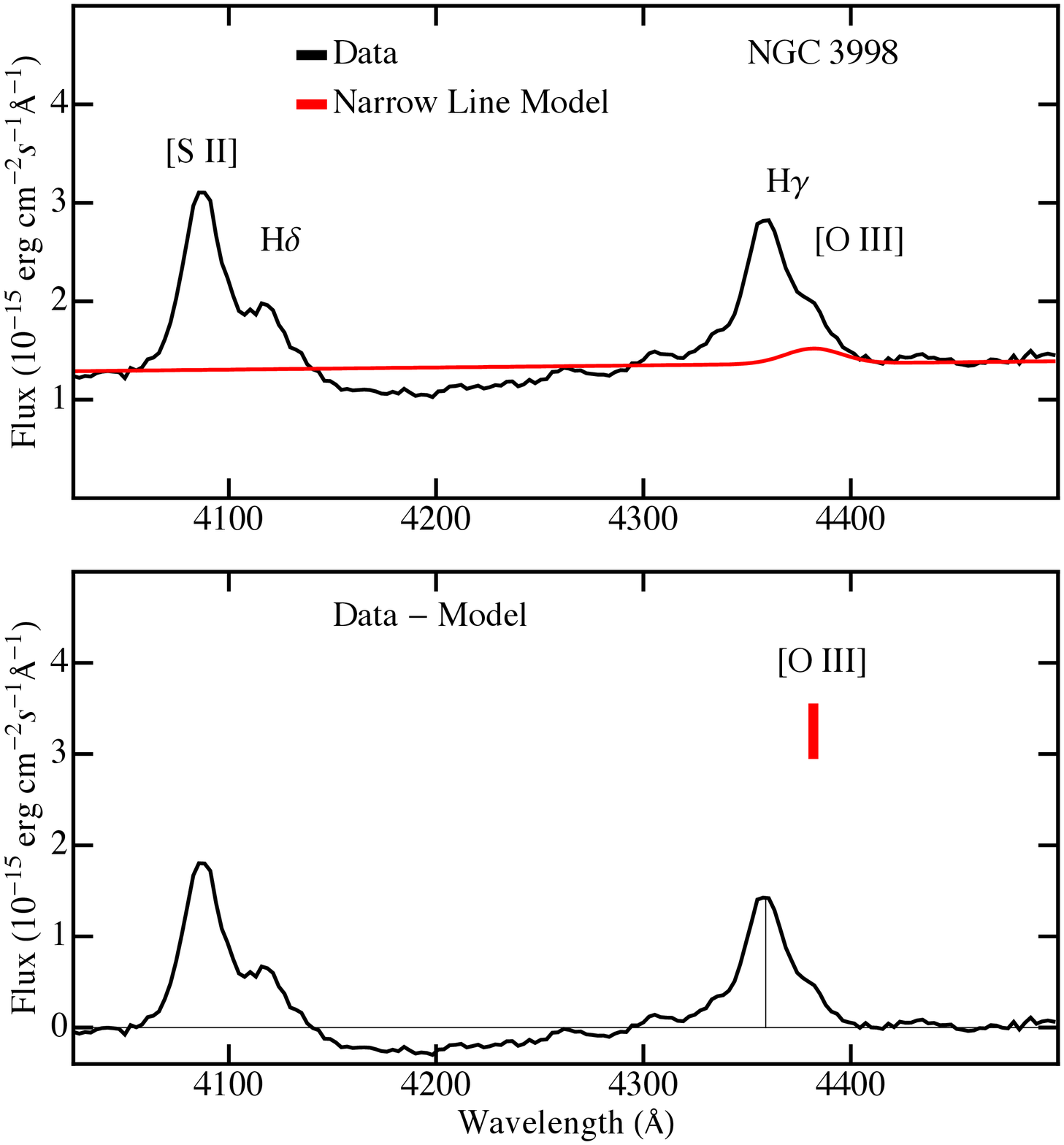}
\caption{{ Broad H${\gamma}$ emission line in NGC 3998. {\sl Top panel}: The observed spectrum
is shown in black and a model for the forbidden line is shown in red (see also Table 2). {\sl Lower panel}: The broad 
H${\gamma}$ emission line profile
after the forbidden line has been subtracted. The central wavelength of the subtracted line is
indicated in red. The vertical black line corresponds to the observed (redshifted) central wavelength of the H${\gamma}$ line }}
\label{default}
\end{center}
\end{figure}

\clearpage

\begin{figure}
\epsscale{0.8}
\begin{center}
\plotone{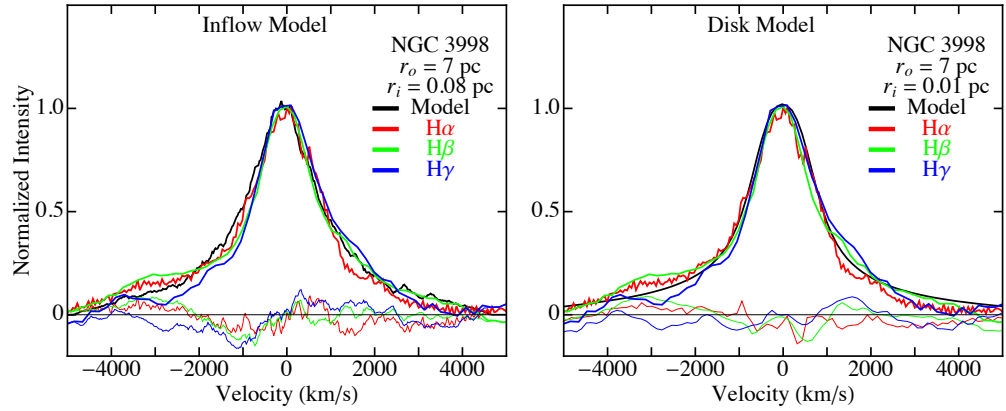}
\caption{{ (Left Panel). Model representation of the broad Balmer line emission in NGC 3998 in terms of a spherically symmetric inflow.  (Right Panel). Model representation of the broad Balmer line emission in NGC 3998 in terms of an accretion disk. The observed 
H${\alpha}$, H${\beta}$ and H${\gamma}$ emission lines are shown in red, green and blue respectively. The model is shown in black. The inner, r${_i}$ and outer radii, r${_o}$ are indicated. Residuals for each profile are plotted as thinner colored lines}}
\label{default}
\end{center}
\end{figure}

\begin{figure}
\epsscale{0.8}
\begin{center}
\plotone{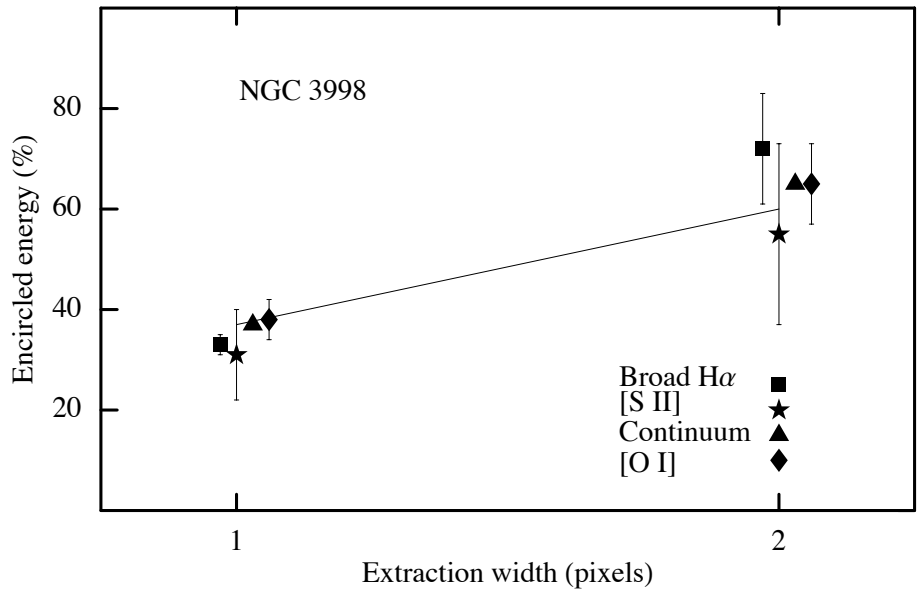}
\caption{{Encircled energy as a function of extraction width. The solid line illustrates the dependence for a point source observed with G750M and a 0.1${\arcsec}$ slit adapted from Fig 13.86 in \cite{Pro10}. Symbols identify the percentage of the flux measured in 1 and 2 pixel extractions as compared with the flux measured in a 7 pixel extraction. The fact that the 
symbols fall near the line indicates that the BELR in NGC 3998 is spatially unresolved with ${\it STIS.}$}}
\label{default}
\end{center}
\end{figure}

\clearpage

\begin{figure}
\epsscale{0.8}
\begin{center}
\plotone{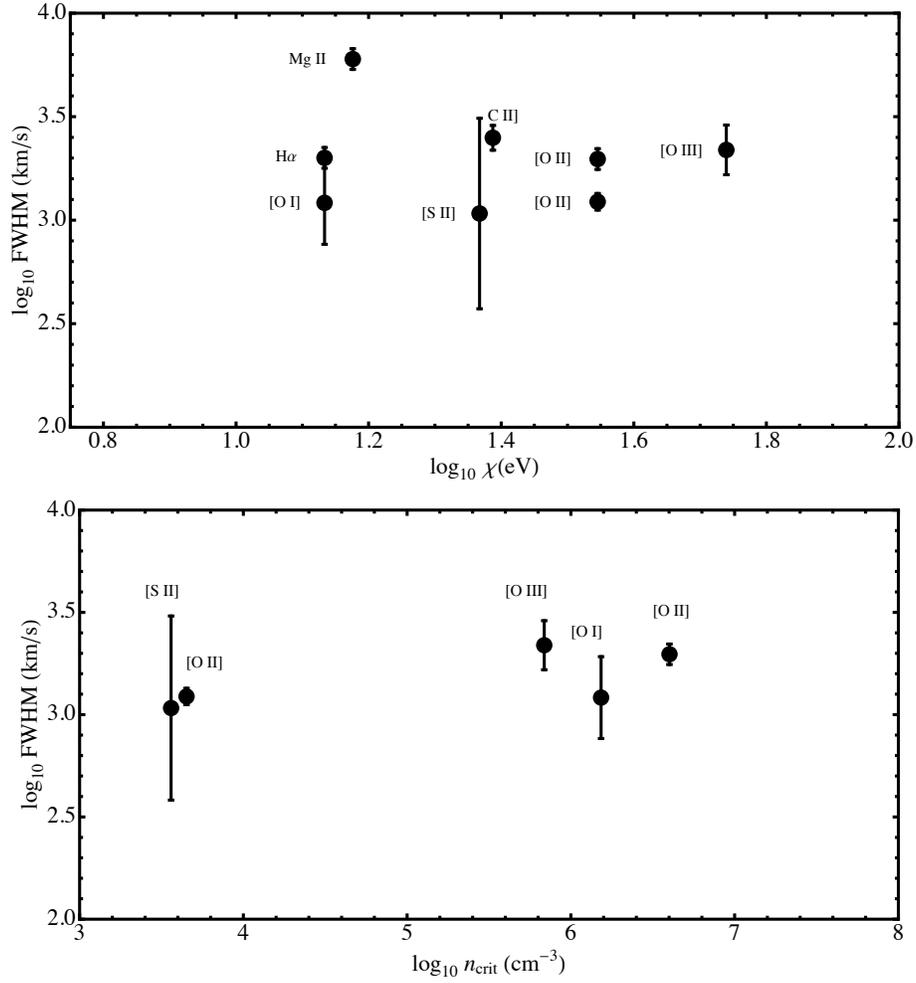}
\caption{{   {\sl Top panel}:  FWHM vs. ionization potential for forbidden and permitted emission lines in NGC 3998. The figure illustrates that the forbidden lines are as broad as the Balmer lines blurring the distinction between the ${narrow}$ and ${broad}$ line regions in this AGN. {\sl Lower panel}: FWHM vs. critical density for forbidden emission lines in NGC 3998.}}
\label{default}
\end{center}
\end{figure}

\clearpage

\begin{deluxetable}{ccccccccc}
\tabletypesize{\scriptsize}
\tablecaption{NGC 3998 Spectral Datasets\label{tbl-2}}
\tablewidth{0pt}
\tablehead{
\colhead{PID} & \colhead{Observation Date} & \colhead{Grating} & \colhead{Spectral Range} & \colhead{Slit} & \colhead{Dispersion} & \colhead{Plate Scale} & \colhead{Integration Time} & \colhead{Datasets}   \\
\colhead{} & \colhead{} &  \colhead{} & \colhead{\AA}  & \colhead{arc sec} & \colhead{\AA/pixel}& \colhead{arc sec/pixel} & \colhead{s} & \colhead{}\\
\colhead{(1)} & \colhead{(2)} &  \colhead{(3)} & \colhead{(4)}  & \colhead{(5)} & \colhead{(6)} & \colhead{(7)} & \colhead{(8)} & \colhead{(9)} \\
}
\startdata
7354 & 11-1-1997 &  G750M  & 6295 - 6867 & 52 x 0.1 & 0.56 & 0.05 & 328 & o4d301030 \\
8839 & 10-2-2000 & G430L & 2900 - 5700 & 52 x 0.2 & 2.73 & 0.05 & 500 & o6a5010a0 \\
8839 & 10-2-2000 & G430L & 2900 - 5700 & 52 x 0.2 & 2.73 & 0.05 & 500 &  o6a501080 \\
8839 & 10-2-2000 & G430L & 2900 - 5700 & 52 x 0.2 & 2.73 & 0.05 & 500 & o6a501090 \\
9486 & 4-7-2002 & G750M &  6482 - 7054    & 52 x 0.1 &  0.56 & 0.05 & 130 & o6n902010 \\
9486 & 4-7-2002 & G750M & 6482 - 7054     & 52 x 0.1 &  0.56 & 0.05 & 130 & o6n902020 \\
9486 & 4-7-2002 & G750M & 6482 - 7054     & 52 x 0.1 &  0.56 & 0.05 & 130 & o6n902030 \\
9486 & 4-7-2002 & G750M &  6482 - 7054    & 52 x 0.1 &  0.56 & 0.05 & 130 &  o6n902040 \\
9486 & 4-7-2002 & G430L & 2900 - 5700 & 52 x 0.2 &  2.73 & 0.05 & 200 & o6n902050 \\
9486 & 4-7-2002 & G430L & 2900 - 5700 & 52 x 0.2 & 2.73 & 0.05 & 200 &  o6n902060 \\
9486 & 4-15-2002 & G140L & 1150 - 1730 & 52 x 0.2 & 0.60 & 0.025 & 2571 & o6n901010 \\
9486 & 4-15-2002 & G230L & 1570 - 3180 & 52 x 0.2 & 1.58 & 0.025 &  3000 & o6n901020 \\
\enddata

\end{deluxetable}

\clearpage

\begin{deluxetable}{cccc}
\tabletypesize{\scriptsize}
\tablecaption{Emission Line Parameters for the G750M Nuclear Spectrum\tablenotemark{a}}
\tablewidth{0pt}
\tablehead{
\colhead{Line} & \colhead{Central Wavelength\tablenotemark{b}} & \colhead{Flux\tablenotemark{c}} & \colhead{FWHM}   \\
\colhead{} & \colhead{\AA} &  \colhead{10$^{-14}$ erg cm$^{-2}$ s$^{-1}$} & \colhead{kms$^{-1}$} \\
\colhead{(1)} & \colhead{(2)} &  \colhead{(3)} & \colhead{(4)} \\
}
\startdata
$\textrm{[O I]}$ &  6325  ${\pm}$ 2 & 7.1 ${\pm}$ 1.0 &  1212 ${\pm}$  244 \\
$\textrm{[O I]}$  &  6388  ${\pm}$ 5 & 2.4 ${\pm}$ 1.1 &  1082 ${\pm}$  612 \\
$\textrm{[N II]}$\tablenotemark{d} &  6576   & ${\leq}$ 2.3  &  1000  \\
H${\alpha}$ (broad)\tablenotemark{e} &6591  & ${\geq}$ 44.7 & 2000 ${\pm}$ 80 \\
H${\alpha}$ (narrow) &6591  & 2.4 & 1000  \\
$\textrm{[N II]}$ & 6612\  & ${\leq}$ 7\tablenotemark{f} & 1000  \\
$\textrm{[S II]}$ & 6741 ${\pm}$ 6  & 1.9 ${\pm}$ 0.8 & 1077 ${\pm}$ 713 \\
$\textrm{[S II]}$ & 6755 ${\pm}$ 3 & 3.3  ${\pm}$ 0.1 & 1077 ${\pm}$ 494 \\
\enddata
\tablenotetext{a}{Table entries that do not include uncertainties are fixed parameters.}
\tablenotetext{b}{Observed wavelength}
\tablenotetext{c}{Measured within
a 0.1{\arcsec}  x 0.35{\arcsec}  aperture. Continuum subtracted but not corrected for dust extinction. }
\tablenotetext{d}{The [NII] 6576 {\AA} emission line flux is constrained by atomic physics to have a flux 1/3 that of the [NII] 6612 line.}
\tablenotetext{e} {The broad H${\alpha}$ emission line flux is an lower limit because the 
[NII] emission lines fluxes that were subtracted are upper limits.}
\tablenotetext{f} {The [NII] emission line flux is an upper limit chosen so as to not over-subtract the broad H${\alpha}$ emission line profile}
\end{deluxetable}

\clearpage

\begin{deluxetable}{cccc}
\tabletypesize{\scriptsize}
\tablecaption{Emission Line Parameters for the G430L Nuclear Spectrum\tablenotemark{a}}
\tablewidth{0pt}
\tablehead{
\colhead{Line} & \colhead{Central Wavelength\tablenotemark{b}} & \colhead{Flux\tablenotemark{c}} & \colhead{FWHM}   \\
\colhead{} & \colhead{\AA} &  \colhead{10$^{-14}$ erg cm$^{-2}$ s$^{-1}$} & \colhead{kms$^{-1}$} \\
\colhead{(1)} & \colhead{(2)} &  \colhead{(3)} & \colhead{(4)} \\
}
\startdata
$\textrm{[O II]}$& 3741  ${\pm}$ 1 & 6.3 ${\pm}$ 0.1 &  1227 ${\pm}$  23 \\
H${\gamma}$ (broad) & 4359  &  4.6 ${\pm}$ 0.2 & 1750 ${\pm}$ 80 \\
$\textrm{[O III]}$&  4382  & ${\leq}$ 0.5\tablenotemark{d}  &  2185 \\
H${\beta}$ (broad) & 4882  &  13.0 ${\pm}$ 0.2 & 1750 ${\pm}$ 80 \\
$\textrm{[O III]}$&  4975  ${\pm}$ 2 & 3.7 ${\pm}$ 0.3 &  2185  \\
$\textrm{[O III]}$&  5027  ${\pm}$ 1 & 8.4 ${\pm}$ 0.4 &  2185 ${\pm}$  265 \\

\enddata
\tablenotetext{a}{Table entries that do not include uncertainties are fixed parameters.}
\tablenotetext{b}{Observed wavelength}
\tablenotetext{c}{Measured within
a 0.2{\arcsec}  x 0.35{\arcsec}  aperture. Continuum subtracted but not corrected for dust extinction. }
\tablenotetext{d} {The [O III] emission line flux is chosen so as to not over-subtract the broad H${\gamma}$ emission line profile}
\end{deluxetable}

\clearpage

\begin{deluxetable}{cccc}
\tabletypesize{\scriptsize}
\tablecaption{Emission Line Parameters for the G230L Nuclear Spectrum\tablenotemark{a}}
\tablewidth{0pt}
\tablehead{
\colhead{Line} & \colhead{Central Wavelength\tablenotemark{b}} & \colhead{Flux\tablenotemark{c}} & \colhead{FWHM}   \\
\colhead{} & \colhead{\AA} &  \colhead{10$^{-14}$ erg cm$^{-2}$ s$^{-1}$} & \colhead{kms$^{-1}$} \\
\colhead{(1)} & \colhead{(2)} &  \colhead{(3)} & \colhead{(4)} \\
}
\startdata
$\textrm{C II]}$\tablenotemark{d} &  2334.4 ${\pm}$ 0.4 & 9.4 ${\pm}$  0.1 &  2500 ${\pm}$ 145 \\
$\textrm{[O II]}$&  2480.4 ${\pm}$ 0.4 & 1.7 ${\pm}$  0.1 &  1974 ${\pm}$ 104 \\
Mg II (broad)\tablenotemark{d} & 2809  &  ${\geq}$18 & ${\sim}$ 6000 \\

\enddata
\tablenotetext{a}{Table entries that do not include uncertainties are fixed parameters.}
\tablenotetext{b}{Observed wavelength}
\tablenotetext{c}{Measured within a 0.2${\arcsec}$  x 0.175${\arcsec}$  aperture. Continuum subtracted but not corrected for dust extinction.}
\tablenotetext{d}{~Line is asymmetric.}

\end{deluxetable}

\clearpage

\clearpage

\begin{deluxetable}{lc}
\tabletypesize{\scriptsize}
\tablewidth{0in}
\tablecaption{Axisymmetric Disk Model Parameters}
\tablehead{
\colhead{Model Parameter} &
\colhead{Value}
}
\startdata
Inclination, $i$            &  45$^{\circ}$ \\
Inner radius, $\xi_1$               &   1,000 $r_{\rm g}\;$\tablenotemark{a} \\
Outer radius, $\xi_1$               & 535,000 $r_{\rm g}\;$\tablenotemark{a} \\
Broadening parameter, $\sigma$      & 300 km s$^{-1}$ \\
Emissivity index, q & 2.0 \\
\enddata
\tablenotetext{a}{Radii are expressed in units of the gravitational
radius, $r_{\rm g}\equiv GM{_{\bullet}}/c^2$, where M${_{\bullet}}$ is the
mass of the central black hole.}
\end{deluxetable}

\end{document}